\documentclass[showpacs,aps,graphicx]{revtex4}
\usepackage{graphicx}

\begin{document}

\title{Efficient entanglement concentration for arbitrary less-entangled N-atom state}

\author{Lan Zhou$^{1,2}$ }
\address{$^1$ College of Mathematics \& Physics, Nanjing University of Posts and Telecommunications, Nanjing,
210003, China\\
$^2$ Institute of Signal Processing  Transmission, Nanjing
University of Posts and Telecommunications, Nanjing, 210003,  China\\}

\date{\today}

\begin{abstract}
A recent paper (Phys. Rev. A \textbf{86}, 034305 (2012)) proposed an entanglement concentration protocol (ECP) for less-entangled $N$-atom GHZ state with the help of the photonic Faraday rotation. It is shown that the maximally entangled atom state can be distilled from two pairs of less-entangled atom states. In this paper, we put forward an improved ECP for arbitrary less-entangled N-atom GHZ state with only one pair of less-entangled atom state, one auxiliary atom and one
auxiliary photon. Moreover, our ECP can be used repeatedly to obtain a higher success probability. If consider the practical operation and imperfect detection, our protocol is more efficient. This ECP may be useful in current quantum information processing.
\end{abstract}

 \maketitle
\section{Introduction}
    In recent years, entanglement has been regarded as a key source in the tasks
of quantum information processing (QIP), for it can hold the power for the quantum nonlocality \cite{Einstein} and provide wide applications \cite{book,rmp}. In all applications, the ideal entangled state is the maximally entangled state. For example, in the quantum teleportation \cite{teleportation,cteleportation}, quantum dense coding \cite{densecoding}, quantum communication \cite{QSDC,QSDC1,QSDC2}, and entanglement-based quantum key distribution \cite{Ekert91,QKDdeng1,QKDdeng2}, one needs to use the maximally entangled state to set up the quantum entanglement channel. Unfortunately, in the practical process, the maximally entangled state may inevitably interact with the channel noise from the environment, which can make the maximally entangled state degrade to the mixed state or pure less-entangled state. In the application process, the mixed state and pure less-entangled state may decrease after the entanglement swapping and cannot ultimately set up the high quality quantum entanglement channel \cite{memory}, so that we need to recover the mixed state or pure less-entangled state into the maximally entangled state.

    Entanglement concentration, which will be detailed here, is a powerful way to recover the pure less-entangled state into the maximally entangled
  state probabilistically \cite{C.H.Bennett2,swapping1,swapping2,zhao1,Yamamoto1,wangxb,shengpra2,shengpra3,shengpra4,dengpra,shengqic,shengpla,wangchuan,wangchuan2,Pengpra}. In 1996, Bennett \emph{et al.} proposed the first entanglement concentration protocol (ECP), which is called the Schmidt projection method \cite{C.H.Bennett2}. Since then, various ECPs have been put forward successively, such as the ECP based on entanglement swapping \cite{swapping1}, the ECP based on the unitary transformation \cite{swapping2}. In 2001, Zhao \emph{et al.} and Yamamoto \emph{et al.} proposed two similar concentration protocols independently with linear optical elements \cite{zhao1,Yamamoto1}, which was later developed by Sheng \emph{et al.} with the help of the cross-Kerr nonlinearity \cite{shengpra2,shengpra3}. So far, most ECPs have focused on photon state, for photon is the best candidate for optimal transmission. Actually, the solid atom is also a good candidate for quantum communication and computation. During the past decade, the cavity quantum electrodynamics (QED) have become a powerful platform for the QIP of the photon-atom states, due to the controllable interaction between atoms and photons \cite{cavity,cavity1,cavity2,cavity3,cavity4,cavity5,cavity6,cavity7,cavity8,cavity9}. Especially, many researchers have showed that with the atoms strongly interacting with local high-quality (Q) cavities, the spatially separated cavities could serve as quantum nodes, and construct a quantum network assisted by the photons acting as a quantum bus \cite{cavity6,node,node1,node2}. However, under current experimental conditions, it is quite difficult to build the high-Q cavity and construct the strong coupling to the confined atoms. Moreover, as the high-Q cavity has to be well isolated from the environment, it seems that the high-Q cavity is unsuitable for efficiently accomplishing the input-output process of the photons.

    Recently, available techniques have achieved the input-output process relevant to optical low-Q cavities, such
as the microtoroidal resonator (MTR) \cite{MTR}. It is attractive and applicable to combine the input-output
process with low-Q cavities, for if achieved, it can accomplish high-quality QIP tasks with currently available techniques. In 2009, An \emph{et al.} put forward an innovative scheme to implement the QIP tasks by moderate cavity-atom coupling with low-Q cavities \cite{r}.
They have shown that when a photon interacts with an atom trapped in a low-Q cavity, different polarization of the input photon can cause different phase rotation on the output photon, which is called the photonic Faraday rotation. The photonic Faraday rotation has attracted great attention, for it only works in low-Q cavities and is insensitive to both cavity decay and atomic spontaneous emission.
  Based on the
Faraday rotation, the protocols for entanglement generation \cite{entanglementgeneration},  quantum teleportation \cite{fengmang3}, controlled teleportation \cite{cteleportationqip}, quantum logic gates \cite{logicgate1}, and entanglement swapping \cite{swapping} were proposed.
In 2012, Peng \emph{et al.} proposed an ECP for the less-entangled atom state with the help of the photonic Faraday rotation \cite{Pengpra}. In the ECP, they can successfully distill one pair of maximally entangled N-atom Greenberger-Horne-Zeilinger (GHZ) state from two pairs of pure  less-entangled N-atom GHZ states with some probability. However, this ECP is not optimal, for two pairs of less-entangled atom state are not necessary. In this paper, we will put forward an improved ECP for less-entangled N-atom GHZ state. In our protocol, we only require a pair of less-entangled atom state and an auxiliary atom, all of which are trapped in the low-Q cavities. With the help of the photonic Faraday rotation, we can successfully distill the maximally entangled atom state with the same success probability as Ref. \cite{Pengpra}. Moreover, our ECP can be used repeatedly to further concentrate the discarded items in Ref. \cite{Pengpra} and obtain a higher success probability. Especially, if we consider the practical operation and imperfect detection, our ECP is more powerful.

This paper is organized as follows: In Sec. II, we first explain the basic principle of the photonic Faraday rotation. In Sec. III, we explain our ECP for the less-entangled two-atom state. In Sec. IV, we extend this ECP to concentrate the $N$-atom GHZ state, which shows that this ECP is more convenient in practical experiment. In Sec. V, we make a discussion and summary.

\section{photonic Faraday rotation}
\begin{figure}[!h]%[tpb]
\begin{center}
\includegraphics[width=8cm,angle=0]{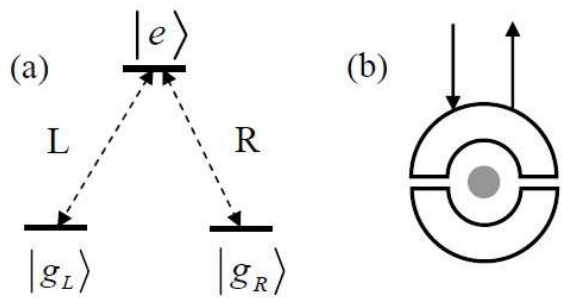}
\caption{A schematic drawing of the interaction between the photon pulse and the three-level atom in the low-Q cavity. (a) the three-level atom trapped in the low-Q cavity. $|g_{L}\rangle$ and $|g_{R}\rangle$ represent two Zeeman sublevels of its degenerate ground state, and $|e\rangle$ represents its excited state. (b) The interaction between the photon pulse and the three-level atom. The state $|g_{L}\rangle$ and $|g_{R}\rangle$ couple with a left (L) polarized and a right (R) polarized photon, respectively.}
\end{center}
\end{figure}

As the photonic Faraday rotation is the key operation in our ECP, before explaining the details of our ECP, we introduce the basic principle of the photonic Faraday rotation firstly. In this section, we present a basic input-output relation for a single photon pulse coherently interacting with a trapped three-level atom. As shown in Fig. 1(a), we suppose that a three-level atom is trapped in the low-Q cavity (one side), where the states $|g_{L}\rangle$ and $|g_{R}\rangle$ represent two Zeeman sublevels of its degenerate ground state, and $|e\rangle$ represents its excited state. A single photon pulse with the frequency $\omega_{p}$ enters the cavity and reacts with the three-level atom. The single photon state can be written as
\begin{eqnarray}
|\varphi_{in}\rangle=\frac{1}{\sqrt{2}}(|L\rangle+|R\rangle),\label{input}
\end{eqnarray}
 where $|L\rangle$ and $|R\rangle$ represent the left-circularly polarization and right-circularly polarization, respectively. The interaction between the photon pulse and the three-level atom can lead the transitions $|g_{L}\rangle\leftrightarrow|e\rangle$ (or $|g_{R}\rangle\leftrightarrow|e\rangle$) for the atom absorbing or
emitting a $|L\rangle$ (or $|R\rangle$) circularly polarized photon. Based on the research from Refs. \cite{Faraday,faraday3,faraday1}, the Hamiltonian of the whole system can be described as
\begin{eqnarray}
H=H_{0}+\hbar\lambda\sum_{j=L,R}(a^{\dagger}_{j}\sigma_{j-}+a_{j}\sigma_{j+})+H_{R},\label{hamiltonian}
\end{eqnarray}
with
\begin{eqnarray}
H_{0}=\sum_{j=L,R}[\frac{\hbar\omega_{0}}{2}\sigma_{jz}+\hbar\omega_{c}a^{\dagger}_{j}a_{j}],\label{h0}
\end{eqnarray}
and
\begin{eqnarray}
H_{R}&=&H_{R0}+i\hbar[\int_{-\infty}^{\infty}d\omega\sum_{j=L,R}\alpha(\omega)(b^{\dagger}_{j}(\omega)a_{j}+b_{j}(\omega)a^{\dagger}_{j})\nonumber\\
&+&\int_{-\infty}^{\infty}d\omega\sum_{j=L,R}\bar{\alpha}(\omega)(c^{\dagger}_{j}(\omega)\sigma_{j-}+c_{j}(\omega)\sigma_{j+})].\label{hr}
\end{eqnarray}
Here, $\lambda$ is the the atom-field coupling constant, $a^{\dagger}_{j}$ and $a_{j}$ are the creation and annihilation operators of the filed-mode in the cavity, respectively, with $j=L, R$. $\sigma_{L-}$ and $\sigma_{L+}$ ($\sigma_{R-}$ and $\sigma_{R+}$) are the lowering and raising operators of the transition L (R), respectively, and $\omega_{c}(\omega_{0})$ is the atomic (field) frequency. In Eq. (\ref{hr}), $H_{R0}$ represents the Hamiltonian of the free reservoirs, and $b_{j}$ and $c_{j}$ ($b^{\dagger}_{j}$ and $c^{\dagger}_{j}$) are the annihilation (creation) operators of the reservoirs.

 Considering the low-Q cavity limit and the
weak excitation limit, we can solve the Langevin equations of
motion for cavity and atomic lowering operators analytically and obtain a single relation between the input and output single-photon state in
the form\cite{r}
\begin{eqnarray}
r(\omega_{p})\equiv\frac{a_{out,j(t)}}{a_{in,j(t)}}=\frac{[i(\omega_{c}-\omega_{p})-\frac{\kappa}{2}][i(\omega_{0}-\omega_{p})+\frac{\gamma}{2}]+g^{2}}
{[i(\omega_{c}-\omega_{p})+\frac{\kappa}{2}][i(\omega_{0}-\omega_{p})+\frac{\gamma}{2}]+g^{2}},\label{r}
\end{eqnarray}
where $\kappa$ and $\gamma$ are the cavity damping rate and atomic decay
rate, respectively, and $g$ is the atom-cavity coupling strength. Eq. (\ref{r}) is a general expression. In the case of the atom uncoupled to the cavity, which makes $g=0$, Eq. (\ref{r}) will simplify as
\begin{eqnarray}
r_{0}(\omega_{p})=\frac{i(\omega_{c}-\omega_{p})-\frac{\kappa}{2}}{i(\omega_{c}-\omega_{p})+\frac{\kappa}{2}}.\label{r0}
\end{eqnarray}

It is obvious that Eq. (\ref{r0}) can be written as a pure phase shift as $r_{0}(\omega_{p})=e^{i\phi_{0}}$. On the other hand, during the interaction process in the cavity, the photon experiences an extremely weak absorption, so that
we can consider that the output photon only experiences a pure phase shift without any absorption for a good approximation. In this case, with strong $\kappa$, weak $\gamma$ and $g$, Eq. (\ref{r}) can be rewritten as $r(\omega_{p})\simeq e^{i\phi}$. In this way, if the photon pulse takes action, the output photon state will convert to $|\varphi_{out}\rangle=r(\omega_{p})|L (R)\rangle\simeq e^{i\phi}|L (R)\rangle$, otherwise, the single-photon pulse would only sense the empty cavity, and the output photon state will convert to $|\varphi_{out}\rangle=r_{0}(\omega_{p})|L (R)\rangle= e^{i\phi_{0}}|L (R)\rangle$. Therefore, for an input single-photon state as Eq. (\ref{input}), if the initial atom state is $|g_{L}\rangle$, the output photon state can be described as
\begin{eqnarray}
|\varphi_{out}\rangle_{-}=\frac{1}{\sqrt{2}}(e^{i\phi}|L\rangle+e^{i\phi_{0}}|R\rangle),\label{L}
\end{eqnarray}
while if the initial atom state is $|g_{R}\rangle$, the output photon state is
\begin{eqnarray}
|\varphi_{out}\rangle_{+}=\frac{1}{\sqrt{2}}(e^{i\phi_{0}}|L\rangle+e^{i\phi}|R\rangle).\label{R}
\end{eqnarray}
Finally, it can be found the polarization direction of the output photon rotates an angle as $\Theta^{-}_{F}=\frac{\phi_{0}-\phi}{2}$ or $\Theta^{+}_{F}=\frac{\phi-\phi_{0}}{2}$, which is called as the photonic Faraday rotation.

Based on the photonic Faraday rotation, it can be seen that in a certain case, i.e., $\omega_{0}=\omega_{c}$, $\omega_{p}=\omega_{c}-\frac{\kappa}{2}$, and $g=\frac{\kappa}{2}$, we can get $\phi=\pi$ and $\phi_{0}=\frac{\pi}{2}$, so that the relation between the input and output photon state can be simplified as \cite{Pengpra}
\begin{eqnarray}
&&|L\rangle|g_{L}\rangle\rightarrow -|L\rangle|g_{L}\rangle,\qquad |R\rangle|g_{L}\rangle\rightarrow i|R\rangle|g_{L}\rangle,\nonumber\\
&&|L\rangle|g_{R}\rangle\rightarrow i|L\rangle|g_{R}\rangle,\qquad |R\rangle|g_{R}\rangle\rightarrow -|R\rangle|g_{R}\rangle.\label{rule}
\end{eqnarray}

Under this special case, the photonic Faraday rotation can be used to perform the entanglement concentration.

\section{The ECP for less-entangled two-atom state}
\begin{figure}[!h]%[tpb]
\begin{center}
\includegraphics[width=8cm,angle=0]{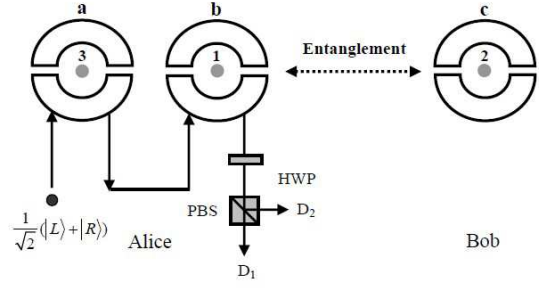}
\caption{A schematic drawing of our ECP for the pure less-entangled two-atom state. The target less-entangled atoms 1 and 2 are trapped in the low-Q cavity b and c, while the auxiliary atom 3 is trapped in the low-Q cavity a. A single-photon pulse passes through the low-Q cavities a and b, and interacts with the atom 3 and 1, successively. By measuring the states of the auxiliary three-level atom and the output photon, we can distill the maximally entangled atom state with some success probability. Moreover, our ECP can be used repeatedly to further concentrate the less-entangled atom state}
\end{center}
\end{figure}

Now we start to explain our ECP for arbitrary less-entangled two-atom state. The schematic drawing of our ECP is shown in Fig. 2. We suppose the two parties Alice and Bob share a pair of less-entangled two-atom state with the form
\begin{eqnarray}
|\psi\rangle_{12}=\alpha|g_{L}g_{R}\rangle_{12}+\beta|g_{R}g_{L}\rangle_{12},\label{initial}
\end{eqnarray}
where the two atoms are marked as 1 and 2, respectively. $\alpha$ and $\beta$ are the initial entanglement coefficients of the atom state, $|\alpha|^{2}+|\beta|^{2}=1$ and $\alpha\neq\beta$. Alice and Bob make the two atoms trapped in the low-Q cavities b and c, respectively. For concentrating the less-entangled atom state, Alice introduces an auxiliary three-level atom 3 with the form
 \begin{eqnarray}
|\psi\rangle_{3}=\beta|g_{L}\rangle_{3}+\alpha|g_{R}\rangle_{3},\label{single atom}
\end{eqnarray}
and makes it trapped in the low-Q cavity a. Here, for preparing the auxiliary single-atom state, we need to know the exact value of $\alpha$ and $\beta$ in advance. Actually, according to the previous research results, we can obtain the exact value of $\alpha$ and $\beta$ by measuring an enough amount of the target samples \cite{swapping2,shengpra2,shengpra3,shengpra4,dengpra}.

Then, Alice makes a single-photon pulse with the form of $|\phi\rangle=\frac{1}{\sqrt{2}}(|L\rangle+|R\rangle)$ pass through the cavities a and b, sequentially. Before entering the cavity, the single photon state combined with the three-level atoms state can be described as
\begin{eqnarray}
|\Psi\rangle&=&|\phi\rangle\otimes|\psi\rangle_{12}\otimes|\psi\rangle_{3}=\alpha\beta|L\rangle|g_{L}g_{R}g_{L}\rangle_{123}
+\alpha\beta|R\rangle|g_{L}g_{R}g_{L}\rangle_{123}\nonumber\\
&+&\beta^{2}|L\rangle|g_{R}g_{L}g_{L}\rangle_{123}
+\beta^{2}|R\rangle|g_{R}g_{L}g_{L}\rangle_{123}
+\alpha^{2}|L\rangle|g_{L}g_{R}g_{R}\rangle_{123}\nonumber\\
&+&\alpha^{2}|R\rangle|g_{L}g_{R}g_{R}\rangle_{123}
+\alpha\beta|L\rangle|g_{R}g_{L}g_{R}\rangle_{123}+\alpha\beta|R\rangle|g_{R}g_{L}g_{R}\rangle_{123}.\label{whole}
\end{eqnarray}

After entering the cavity, the photon interacts with the three-level atom. Based on Eq. (\ref{rule}), we can get the relation between the input and the output photon state when the photon pulse passes through two cavities, sequentially, which can be written as
\begin{eqnarray}
&&|L\rangle|g_{L}g_{L}\rangle\rightarrow |L\rangle|g_{L}g_{L}\rangle,\qquad|R\rangle|g_{L}g_{L}\rangle\rightarrow -|R\rangle|g_{L}g_{L}\rangle,\nonumber\\
&&|L\rangle|g_{L}g_{R}\rangle\rightarrow -i|L\rangle|g_{L}g_{R}\rangle,\qquad|R\rangle|g_{L}g_{R}\rangle\rightarrow -i|R\rangle|g_{L}g_{R}\rangle,\nonumber\\
&&|L\rangle|g_{R}g_{L}\rangle\rightarrow -i|L\rangle|g_{R}g_{L}\rangle,\qquad|R\rangle|g_{R}g_{L}\rangle\rightarrow -i|R\rangle|g_{R}g_{L}\rangle,\nonumber\\
&&|L\rangle|g_{R}g_{R}\rangle\rightarrow -|L\rangle|g_{R}g_{R}\rangle,\qquad|R\rangle|g_{R}g_{R}\rangle\rightarrow |R\rangle|g_{R}g_{R}\rangle.\label{rule2}
\end{eqnarray}
Therefore, after the photon pulse emitting from the low-Q cavity b, Eq. (\ref{whole}) can evolve to
\begin{eqnarray}
|\Psi\rangle&\rightarrow&|\Psi\rangle_{out}=\alpha\beta|L\rangle|g_{L}g_{R}g_{L}\rangle_{123}+(-\alpha\beta)|R\rangle|g_{L}g_{R}g_{L}\rangle_{123}\nonumber\\
&+&(-i\beta^{2})|L\rangle|g_{R}g_{L}g_{L}\rangle_{123}+(-i\beta^{2})|R\rangle|g_{R}g_{L}g_{L}\rangle_{123}
+(-i\alpha^{2})|L\rangle|g_{L}g_{R}g_{R}\rangle_{123}\nonumber\\
&+&(-i\alpha^{2})|R\rangle|g_{L}g_{R}g_{R}\rangle_{123}+(-\alpha\beta)|L\rangle|g_{R}g_{L}g_{R}\rangle_{123}+\alpha\beta|R\rangle|g_{R}g_{L}g_{R}\rangle_{123}.\label{out}
\end{eqnarray}

Then Alice performs the Hadamard operation on the atom 3 by driving atom 3 with an external classical field (polarized lasers), which makes
\begin{eqnarray}
|g_{L}\rangle\rightarrow\frac{1}{\sqrt{2}}(|g_{L}\rangle+|g_{R}\rangle), \qquad |g_{R}\rangle\rightarrow\frac{1}{\sqrt{2}}(|g_{L}\rangle-|g_{R}\rangle).
\end{eqnarray}
After the Hadamard operation, Eq. (\ref{out}) can evolve to
\begin{eqnarray}
|\Psi\rangle_{out}&\rightarrow&(\alpha\beta|L-R\rangle|g_{L}g_{R}\rangle_{12}-i\beta^{2}|L+R\rangle|g_{R}g_{L}\rangle_{12}
-i\alpha^{2}|L+R\rangle|g_{L}g_{R}\rangle_{12}\nonumber\\
&-&\alpha\beta|L-R\rangle|g_{R}g_{L}\rangle_{12})|g_{L}\rangle_{3}+(\alpha\beta|L-R\rangle|g_{L}g_{R}\rangle_{12}
-i\beta^{2}|L+R\rangle|g_{R}g_{L}\rangle_{12}\nonumber\\
&+&i\alpha^{2}|L+R\rangle|g_{L}g_{R}\rangle_{12}+\alpha\beta|L-R\rangle|g_{R}g_{L}\rangle_{12})|g_{R}\rangle_{3}.\label{out1}
\end{eqnarray}

Then, Alice makes output photon pass through a quarter-wave plate (QWP), which can make
\begin{eqnarray}
|L\rangle\rightarrow\frac{1}{\sqrt{2}}(|H\rangle+|V\rangle), \qquad |R\rangle\rightarrow\frac{1}{\sqrt{2}}(|H\rangle-|V\rangle),
\end{eqnarray}
where $|H\rangle$ represents the horizontal polarization and $|V\rangle$ represents the vertical polarization of the photon. After the QWP, Eq. (\ref{out1}) can ultimately evolve to
\begin{eqnarray}
|\Psi_{1}\rangle_{out}&=&\alpha\beta(|g_{L}g_{R}\rangle_{12}-|g_{R}g_{L}\rangle_{12})|V\rangle|g_{L\rangle_{3}}
-(i\alpha^{2}|g_{L}g_{R}\rangle_{12}+i\beta^{2}|g_{R}g_{L}\rangle_{12})|H\rangle|g_{L}\rangle_{3}\nonumber\\
&+&\alpha\beta(|g_{L}g_{R}\rangle_{12}+|g_{R}g_{L}\rangle_{12})|V\rangle|g_{R\rangle_{3}}
+(i\alpha^{2}|g_{L}g_{R}\rangle_{12}-i\beta^{2}|g_{R}g_{L}\rangle_{12})|H\rangle|g_{R}\rangle_{3}.\label{out2}
\end{eqnarray}

Finally, Alice makes the photon enter the PBS, which can transmit then horizontal polarized ($|H\rangle$) photon and reflect the vertical polarization ($|V\rangle$) photon, respectively. After the PBS, both the output photon state and the auxiliary atom state are detected by the detectors. Based on the measurement results, there are four possible cases. If the measurement result is $|V\rangle|g_{R}\rangle_{3}$, Eq. (\ref{out2}) will collapse to
\begin{eqnarray}
|\psi_{1}\rangle_{12}=\frac{1}{\sqrt{2}}(|g_{L}g_{R}\rangle_{12}+|g_{R}g_{L}\rangle_{12}),\label{max}
\end{eqnarray}
while if the measurement result is $|V\rangle|g_{L}\rangle_{3}$, Eq. (\ref{out2}) will collapse to
\begin{eqnarray}
|\psi'_{1}\rangle_{12}=\frac{1}{\sqrt{2}}(|g_{L}g_{R}\rangle_{12}-|g_{R}g_{L}\rangle_{12}).\label{max1}
\end{eqnarray}
It can be found that both the Eq. (\ref{max}) and Eq. (\ref{max1}) are the maximally-entangled atom states, and there is only a phase difference between them. Eq. (\ref{max1}) can be easily converted to Eq. (\ref{max}) by the phase flip operation. So far, we have successfully distilled the maximally-entangled atom state, with the success probability of P=$2|\alpha\beta|^{2}$, which is the same as that in Ref. \cite{Pengpra}. On the other hand, if the measurement result is $|H\rangle|g_{L}\rangle_{3}$, Eq. (\ref{out2}) will collapse to
\begin{eqnarray}
|\psi_{2}\rangle_{12}=\alpha^{2}|g_{L}g_{R}\rangle_{12}+\beta^{2}|g_{R}g_{L}\rangle_{12},\label{new}
\end{eqnarray}
while if the result is $|H\rangle|g_{R}\rangle_{3}$, Eq. (\ref{out2}) will collapse to
\begin{eqnarray}
|\psi'_{2}\rangle_{12}=\alpha^{2}|g_{L}g_{R}\rangle_{12}-\beta^{2}|g_{R}g_{L}\rangle_{12}.\label{new1}
\end{eqnarray}
Similarly, Eq. (\ref{new1}) can be converted to Eq. (\ref{new}) by the phase flip operation. Interestingly, it can be found that Eq. (\ref{new}) has the similar form as Eq. (\ref{initial}), that is to say, Eq. (\ref{new}) is a new less-entangled atom state and can be reconcentrated for the next round. According to the concentration step described above, in the second concentration round, Alice introduces another auxiliary atom 3' with the form
 \begin{eqnarray}
|\psi\rangle_{3'}=\beta^{2}|g_{L}\rangle_{3'}+\alpha^{2}|g_{R}\rangle_{3'},\label{single atom1}
\end{eqnarray}
and also makes it trapped in the low-Q cavity a. By making a single-photon pulse with the form of $|\phi\rangle=\frac{1}{\sqrt{2}}(|L\rangle+|R\rangle)$ pass through the cavities a and b successively, the new less-entangled atom state combined with the single-photon state can evolve to
\begin{eqnarray}
|\Psi'\rangle_{out}&=&\alpha^{2}\beta^{2}|L\rangle|g_{L}g_{R}g_{L}\rangle_{123'}+(-\alpha^{2}\beta^{2})|R\rangle|g_{L}g_{R}g_{L}\rangle_{123'}
+(-i\beta^{4})|L\rangle|g_{R}g_{L}g_{L}\rangle_{123'}\nonumber\\
&+&(-i\beta^{4})|R\rangle|g_{R}g_{L}g_{L}\rangle_{123'}
+(-i\alpha^{4})|L\rangle|g_{L}g_{R}g_{R}\rangle_{123'}
+(-i\alpha^{4})|R\rangle|g_{L}g_{R}g_{R}\rangle_{123'}\nonumber\\
&+&(-\alpha^{2}\beta^{2})|L\rangle|g_{R}g_{L}g_{R}\rangle_{123'}+\alpha^{2}\beta^{2}|R\rangle|g_{R}g_{L}g_{R}\rangle_{123'}.\label{out3}
\end{eqnarray}

Then, Alice performs the Hadamard operation on the atom 3' and the single photon, respectively, and the Eq. (\ref{out3}) can ultimately evolve to
\begin{eqnarray}
|\Psi'_{1}\rangle_{out}&=&\alpha^{2}\beta^{2}(|g_{L}g_{R}\rangle_{12}-|g_{R}g_{L}\rangle_{12})|V\rangle|g_{L}\rangle_{3'}
-(i\alpha^{4}|g_{L}g_{R}\rangle_{12}+i\beta^{4}|g_{R}g_{L}\rangle_{12})|H\rangle|g_{L}\rangle_{3'}\nonumber\\
&+&\alpha^{2}\beta^{2}(|g_{L}g_{R}\rangle_{12}+|g_{R}g_{L}\rangle_{12})|V\rangle|g_{R}\rangle_{3'}
+(i\alpha^{4}|g_{L}g_{R}\rangle_{12}-i\beta^{4}|g_{R}g_{L}\rangle_{12})|H\rangle|g_{R}\rangle_{3'}.\label{out4}
\end{eqnarray}

By measuring the quantum states of the auxiliary atom and the output photon, it can be found that if the measurement result is $|V\rangle|g_{R}\rangle_{3'}$, Eq. (\ref{out4}) will collapse to Eq. (\ref{max}), while if the measurement result is $|V\rangle|g_{L}\rangle_{3'}$, Eq. (\ref{out4}) will collapse to Eq. (\ref{max1}). Therefore, in the second concentration round, we can successfully distill the maximally entangled atom state with the probability P$_{2}=\frac{2|\alpha\beta|^{4}}{|\alpha|^{4}+|\beta|^{4}}$, where the subscript 2' means in the second concentration round. On the other hand, if the measurement result is $|H\rangle|g_{L}\rangle_{3'}$, Eq. (\ref{out4}) will collapse to
\begin{eqnarray}
|\psi_{3}\rangle_{12}=\alpha^{4}|g_{L}g_{R}\rangle_{12}+\beta^{4}|g_{R}g_{L}\rangle_{12},\label{new2}
\end{eqnarray}
while if the measurement result is $|H\rangle|g_{R}\rangle_{3'}$, Eq. (\ref{out4}) will collapse to
\begin{eqnarray}
|\psi'_{3}\rangle_{12}=\alpha^{4}|g_{L}g_{R}\rangle_{12}-\beta^{4}|g_{R}g_{L}\rangle_{12}.\label{new3}
\end{eqnarray}
Eq. (\ref{new3}) can be converted to Eq. (\ref{new2}) by the phase flip operation. Similar to Eq. (\ref{new}),  Eq. (\ref{new2}) is a new less-entangled atom state and can be reconcentrated for the third round. Therefore, our ECP can be used repeatedly to further concentrate the less-entangled two-atom state.

\section{The ECP for less-entangled N-atom GHZ state}
\begin{figure}[!h]%[tpb]
\begin{center}
\includegraphics[width=8cm,angle=0]{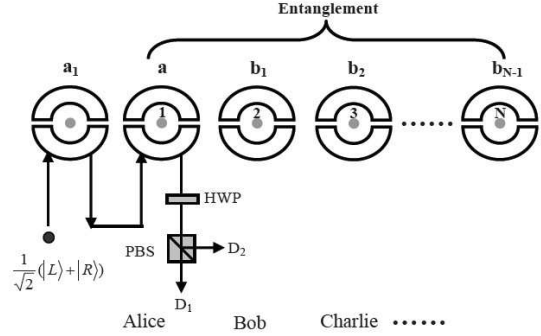}
\caption{A schematic drawing of our ECP for the pure less-entangled N-atom GHZ state. The less-entangled N atoms are trapped in N low-Q cavities with the name of a, b$_{1}$, b$_{2}$, $\cdots$ b$_{N-1}$, which are hold by N parties, respectively. Alice introduce an auxiliary single atom and makes it trapped in the low-Q cavity a$_{1}$ in his hand. Alice makes a single-photon pulse pass through the low-Q cavities a$_{1}$ and a, successively. After the interaction between the photon and the atoms, by measuring the states of the auxiliary atom and the output photon, we can successfully distill the maximally entangled N-atom state. Our ECP can also be used repeatedly to further concentrate the less-entangled N-atom GHZ state.}
\end{center}
\end{figure}

Our ECP can be extended to concentrate the less-entangled N-atom GHZ state. We suppose N entangled three-level atoms which marked as 1, 2, $\cdots$, N are possessed by N parties, say Alice, Bob, Charlie and so on, respectively. Each of the N parties makes his atom trapped in a low-Q cavity in his position. Here, the low-Q cavity in Alice's position is marked as a, while other $N-1$ low-Q cavities in other parties' position are marked as b$_{1}$, $\cdots$, b$_{N-1}$, respectively. In this way, we suppose the less-entangled N-atom GHZ state can be written as
\begin{eqnarray}
|\psi\rangle_{N}=\alpha|g_{L}g_{R}\cdots g_{R}\rangle_{ab_{1}\cdots b_{N-1}}+\beta|g_{R}g_{L}\cdots g_{L}\rangle_{ab_{1}\cdots b_{N-1}}.\label{N atom}  \end{eqnarray}

 Similarly, Alice prepares an auxiliary single atom with the form as Eq. (\ref{single atom}) and makes it trapped in another low-Q cavity a$_{1}$ in his position. Then, Alice makes a single-photon pulse with the form $|\phi\rangle=\frac{1}{\sqrt{2}}(|L\rangle+|R\rangle)$ pass through the cavity a$_{1}$ and a, successively. According to the relationship between the input and output photon in Eq. (\ref{rule2}), After the photon pass through the two cavities, the output photon combined with the whole $N+1$ atom state can evolve to
\begin{eqnarray}
&&|\phi\rangle\otimes|\psi\rangle_{a_{1}}\otimes|\psi\rangle_{N}\rightarrow|\Psi\rangle_{outN}=\alpha\beta|L\rangle|g_{L}g_{L}\rangle_{a_{1}a}|g_{R}\cdots g_{R}\rangle_{b_{1}\cdots b_{N-1}}+(-\alpha\beta)|R\rangle|g_{L}g_{L}\rangle_{a_{1}a}|g_{R}\cdots g_{R}\rangle_{b_{1}\cdots b_{N-1}}\nonumber\\
&+&(-\alpha\beta)|L\rangle|g_{R}g_{R}\rangle_{a_{1}a}|g_{L}\cdots g_{L}\rangle_{b_{1}\cdots b_{N-1}}+\alpha\beta|R\rangle|g_{R}g_{R}\rangle_{a_{1}a}|g_{L}\cdots g_{L}\rangle_{b_{1}\cdots b_{N-1}}\nonumber\\
&+&(-i\alpha^{2})|L\rangle|g_{R}g_{L}\rangle_{a_{1}a}|g_{R}\cdots g_{R}\rangle_{b_{1}\cdots b_{N-1}}+(-i\alpha^{2})|R\rangle|g_{R}g_{L}\rangle_{a_{1}a}|g_{R}\cdots g_{R}\rangle_{b_{1}\cdots b_{N-1}}\nonumber\\
&+&(-i\beta^{2})|L\rangle|g_{L}g_{R}\rangle_{a_{1}a}|g_{L}\cdots g_{L}\rangle_{b_{1}\cdots b_{N-1}}+(-i\beta^{2})|R\rangle|g_{L}g_{R}\rangle_{a_{1}a}|g_{L}\cdots g_{L}\rangle_{b_{1}\cdots b_{N-1}}.\label{outN}
\end{eqnarray}

Then, Alice performs the Hadamard operation on the auxiliary single atom and Eq. (\ref{outN}) will convert to
\begin{eqnarray}
|\Psi\rangle_{outN}&\rightarrow&(\alpha\beta|L-R\rangle|g_{L}g_{R}\cdots g_{R}\rangle_{ab_{1}\cdots b_{N-1}}-\alpha\beta|L-R\rangle|g_{R}g_{L}\cdots g_{L}\rangle_{ab_{1}\cdots b_{N-1}}\nonumber\\
&-&i\alpha^{2}|L+R\rangle|g_{L}g_{R}\cdots g_{R}\rangle_{ab_{1}\cdots b_{N-1}}-i\beta^{2}|L+R\rangle|g_{R}g_{L}\cdots g_{L}\rangle_{ab_{1}\cdots b_{N-1}})|g_{L}\rangle\nonumber\\
&+&(\alpha\beta|L-R\rangle|g_{L}g_{R}\cdots g_{R}\rangle_{ab_{1}\cdots b_{N-1}}+\alpha\beta|L-R\rangle|g_{R}g_{L}\cdots g_{L}\rangle_{ab_{1}\cdots b_{N-1}}\nonumber\\
&+&i\alpha^{2}|L+R\rangle|g_{L}g_{R}\cdots g_{R}\rangle_{ab_{1}\cdots b_{N-1}}-i\beta^{2}|L+R\rangle|g_{R}g_{L}\cdots g_{L}\rangle_{ab_{1}\cdots b_{N-1}})|g_{R}\rangle.
\end{eqnarray}

Alice makes the output photon pass through the HWP and the PBS, successively. After the PBS,  Eq. (\ref{outN}) can finally evolve to
\begin{eqnarray}
|\Psi_{1}\rangle_{outN}&=&\alpha\beta(|g_{L}g_{R}\cdots g_{R}\rangle_{ab_{1}\cdots b_{N-1}}-|g_{R}g_{L}\cdots g_{L}\rangle_{ab_{1}\cdots b_{N-1}})|V\rangle|g_{L}\rangle\nonumber\\
&-&(i\alpha^{2}|g_{L}g_{R}\cdots g_{R}\rangle_{ab_{1}\cdots b_{N-1}}+i\beta^{2}|g_{L}g_{R}\cdots g_{R}\rangle_{ab_{1}\cdots b_{N-1}})|H\rangle|g_{L}\rangle\nonumber\\
&+&\alpha\beta(|g_{L}g_{R}\cdots g_{R}\rangle_{ab_{1}\cdots b_{N-1}}+|g_{R}g_{L}\cdots g_{L}\rangle_{ab_{1}\cdots b_{N-1}})|V\rangle|g_{R}\rangle\nonumber\\
&+&(i\alpha^{2}|g_{L}g_{R}\cdots g_{R}\rangle_{ab_{1}\cdots b_{N-1}}-i\beta^{2}|g_{L}g_{R}\cdots g_{R}\rangle_{ab_{1}\cdots b_{N-1}})|H\rangle|g_{H}\rangle.\label{outN1}
\end{eqnarray}

Finally, Alice measures the state of the auxiliary atom and the output photon. It can be found that there are still four possible cases based on different measurement results. If the result is $|V\rangle|g_{R}\rangle$
, Eq. (\ref{outN1}) will collapse to
\begin{eqnarray}
|\psi_{1}\rangle_{N}=\frac{1}{\sqrt{2}}(|g_{L}g_{R}\cdots g_{R}\rangle_{ab_{1}\cdots b_{N-1}}+|g_{R}g_{L}\cdots g_{L}\rangle_{ab_{1}\cdots b_{N-1}}),\label{maxN}
\end{eqnarray}
while if the result is $|V\rangle|g_{L}\rangle$, Eq. (\ref{outN1}) will collapse to
\begin{eqnarray}
|\psi'_{1}\rangle_{N}=\frac{1}{\sqrt{2}}(|g_{L}g_{R}\cdots g_{R}\rangle_{ab_{1}\cdots b_{N-1}}-|g_{R}g_{L}\cdots g_{L}\rangle_{ab_{1}\cdots b_{N-1}}).\label{maxN1}
\end{eqnarray}
 Both Eq. (\ref{maxN}) and Eq. (\ref{maxN1}) are the maximally entangled N-atom GHZ state. Eq. (\ref{maxN1}) can be converted to Eq. (\ref{maxN}) easily by the phase flip operation. Therefore, so far, we successfully distill the maximally entangled N-atom GHZ state from the less-entangled N-atom state, with the success probability of P=$2|\alpha\beta|^{2}$.

On the other hand, if the measurement result is $|H\rangle|g_{L}\rangle$, Eq. (\ref{outN1}) will collapse to
\begin{eqnarray}
|\psi_{2}\rangle_{N}=\alpha^{2}|g_{L}g_{R}\cdots g_{R}\rangle_{ab_{1}\cdots b_{N-1}}+\beta^{2}|g_{L}g_{R}\cdots g_{R}\rangle_{ab_{1}\cdots b_{N-1}},\label{new4}
\end{eqnarray}
while if the result is $|H\rangle|g_{R}\rangle$, Eq. (\ref{outN1}) will collapse to
\begin{eqnarray}
|\psi'_{2}\rangle_{N}=\alpha^{2}|g_{L}g_{R}\cdots g_{R}\rangle_{ab_{1}\cdots b_{N-1}}-\beta^{2}|g_{L}g_{R}\cdots g_{R}\rangle_{ab_{1}\cdots b_{N-1}}.\label{new5}
\end{eqnarray}
Eq. (\ref{new5}) can be converted to Eq. (\ref{new4}) by the phase flip operation. Similar to the Sec.III, it can be seen that Eq. (\ref{new4}) is a new less-entangled N-atom GHZ state. Based on the concentration step described above, Alice only needs to prepare a new auxiliary single three-level atom with the form
\begin{eqnarray}
|\psi_{1}\rangle_{a_{1}}=\beta^{2}|g_{L}\rangle_{a_{1}}+\alpha^{2}|g_{R}\rangle_{a_{1}},
\end{eqnarray}
and Eq. (\ref{new5}) can be reconcentrated for the next round. Therefore, it has shown that our ECP can be used repeatedly to further concentrate the less-entangled N-atom state and the success probability in each concentration round is the same as that in Sec. III.

\section{discussion and summary}
\begin{figure}[!h]%[tpb]
\begin{center}
\includegraphics[width=8cm,angle=0]{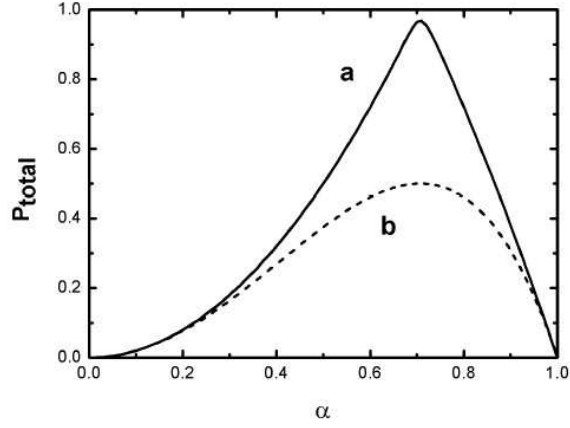}
\caption{The success probability ($P_{total}$) of both ECPs for obtaining a maximally entangled N-atom GHZ state entangled state under ideal conditions. (a) The $P_{total}$ after our ECP being operated for K times. For numerical simulation, we choose $K=5$. (b) The $P_{total}$ of the ECP in Ref. \cite{Pengpra}. It can be seen that the value of $P_{total}$ largely depends on the initial coefficient $\alpha$. Moreover, by repeating our ECP for 5 times, the $P_{total}$ of our ECP is much higher than that in Ref. \cite{Pengpra}.}
\end{center}
\end{figure}
 It is interesting to calculate the total success probability in our ECP. According to the concentration step in Sec. III and Sec. IV, we can calculate the success probability of our ECP in each concentration round as
\begin{eqnarray}
P_{1}&=&2|\alpha\beta|^{2},\nonumber\\
P_{2}&=&\frac{2|\alpha\beta|^{4}}{|\alpha|^{4}+|\beta|^{4}},\nonumber\\
P_{3}&=&\frac{2|\alpha\beta|^{8}}{(|\alpha|^{4}+|\beta|^{4})(|\alpha|^{8}+|\beta|^{8})},\nonumber\\
P_{4}&=&\frac{2|\alpha\beta|^{16}}{(|\alpha|^{4}+|\beta|^{4})(|\alpha|^{8}+|\beta|^{8})(|\alpha|^{16}+|\beta|^{16})},\nonumber\\
&\cdots\cdots&\nonumber\\
P_{K}&=&\frac{2|\alpha\beta|^{2^{K}}}{(|\alpha|^{4}+|\beta|^{4})(|\alpha|^{8}+|\beta|^{8})\cdots(|\alpha|^{2^{K}}+|\beta|^{2^{K}})^{2}},\label{probability}
\end{eqnarray}
where the success probability of the ECP in Ref. \cite{Pengpra} only equals the $P_{1}$ of our ECP.

As our ECP can be used indefinitely in theory, the total success probability P$_{total}$ equals the sum of the probability in each concentration round, which can be written as
\begin{eqnarray}
P_{total}=P_{1}+P_{2}+\cdots P_{K}=\sum\limits_{K=1}^{\infty} P_{K}.
\end{eqnarray}
It is obvious that if the original state is the maximally entangled N-atom state, where $\alpha=\beta=\frac{1}{\sqrt{2}}$, the probability $P_{total}=\frac{1}{2}+\frac{1}{4}+\frac{1}{8}+\cdots\frac{1}{2^{K}}+\cdots=1$, while if the original state is the less-entangled N-atom state, where $\alpha\neq\beta$, the $P_{total}<1$. Fig. 4 shows the $P_{total}$ of our ECP and the ECP in Ref. \cite{Pengpra} as a function of the initial entanglement coefficient $\alpha$, where we choose $K=5$ for a proper approximation. It can be found that $P_{total}$ largely depends on the original entanglement state and by repeating the our ECP for 5 times, the $P_{total}$ of our ECP is much higher than that of Ref. \cite{Pengpra}.

In the concentration process, the atom state detection and photon state detection play prominent roles. In Fig. 4, we have assumed that both the two detections are perfect with the detection efficiency $\eta=100\%$. Actually, in current experimental conditions, the detection efficiency $\eta<100\%$. Therefore, it is worthy to compare the success probability of the two ECPs under practical experimental conditions. In the ECP of Ref. \cite{Pengpra}, the output single photon state and a pair of N-atom state need to be detected. We suppose that the single photon detection efficiency and the single atom detection efficiency are $\eta_{p}$ and $\eta_{a}$, respectively. Therefore, the success probability of the ECP in Ref. \cite{Pengpra} can be revised as,
\begin{eqnarray}
P'_{total}&=&\eta_{p}\eta_{a}^N2|\alpha\beta|^{2}\label{p}
\end{eqnarray}
Eq. (\ref{p}) indicates that the success probability of the ECP in Ref. \cite{Pengpra} shows an exponential decay with the atom number N.

\begin{figure}[!h]%[tpb]
\begin{center}
\includegraphics[width=8cm,angle=0]{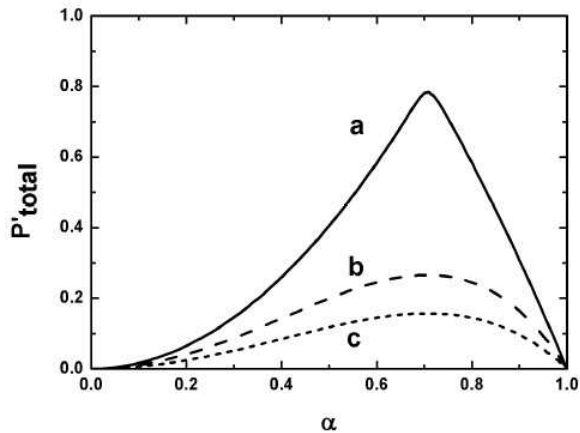}
\caption{ The success probability ($P'_{total}$) for both two ECPs under the imperfect detection, where the
efficiencies of both the photon state detection and atom state detection are set as $90\%$ for approximation. (a) The $P'_{total}$ after our ECP being operated for 5 times. (b) The $P'_{total}$ of the ECP in Ref. \cite{Pengpra}, when the atom number N is 5. (c) The $P'_{total}$ of the ECP in Ref. \cite{Pengpra}, when the atom number N is 10. It can be seen under imperfect detection, the $P'_{total}$ of the ECP in Ref. \cite{Pengpra} reduces largely with the atom number N, while the $P'_{total}$ of our ECP can remains a relatively high level.}
\end{center}
\end{figure}

On the other hand, in our ECP, we only need to detect the output single photon state and the auxiliary single atom state in each concentration round, so that the success probability in each concentration round can be described as
\begin{eqnarray}
P'_{K}=\eta_{p}\eta_{a}P_{K}=\eta_{p}\eta_{a}\frac{2|\alpha\beta|^{2^{K}}}{(|\alpha|^{4}+|\beta|^{4})(|\alpha|^{8}+|\beta|^{8})\cdots(|\alpha|^{2^{K}}+|\beta|^{2^{K}})^{2}}.
\end{eqnarray}
Therefore, the total success probability P$_{total}$ can be revised as
\begin{eqnarray}
P'_{1total}=\eta_{p}\eta_{a}(P_{1}+P_{2}+\cdots P_{K})=\eta_{p}\eta_{a}P_{total},\label{pt}
\end{eqnarray}
which indicates the success probability of our ECP is independent of the atom number N.

Fig. 5 shows the values of $P'_{total}$ and $P'_{1total}$ as a function of the entanglement coefficient $\alpha$. For numerical simulation, we assume $\eta_{p}=90\%$ and $\eta_{a}=90\%$ for approximation. In our ECP, we choose the repeating number $K=5$ (Fig. 5(a)), while in the ECP in Ref. \cite{Pengpra}, we choose the atom number $N=5$ (Fig. 5(b)) and $N=10$ (Fig. 5(b)). It is obvious that $P'_{total}$ of the ECP in Ref. \cite{Pengpra} reduces largely with the increasing of the atom number N. Especially, according to Eq. (\ref{p}) and Eq. (\ref{pt}), when the atom number N is large, the $P'_{total}\rightarrow 0$, while our ECP can still get high success probability. Therefore, under practical experiment conditions, especially when the atom number N is large, our ECP shows greater advantage.

So far, we have fully described our ECP. With the help of the photonic Faraday rotation, we can successfully distill the maximally entangled N-atom state from arbitrary less-entangled N-atom state. Comparing with the ECP from Ref. \cite{Pengpra}, our ECP is more efficient. Firstly, our ECP reduces one pair of less-entangled N-atom GHZ state but obtains the same success probability. Secondly, by repeating our ECP, the discarded items in Ref.\cite{Pengpra} can be reused, so that our ECP can obtain a higher success probability. On the other hand, it is shown that our ECP is more powerful for concentrating the less-entangled multipartite state. In Ref.\cite{Pengpra}, after successfully performed the ECP, each of the N parties should measure his or her atom to ultimately obtain the maximally entangled N-atom GHZ state. Finally, they also should check their measurement results to confirm the remained maximally entangled state. Therefore, it increases the operating complexity greatly, especially when the atom number N is large.  In our ECP, Alice can operate the whole protocol alone. After the concentration, she only needs to tell others the final results, which can greatly reduce the practical operations. Moreover, according to above discussion, if we consider the imperfect detection, the success probability of the ECP in Ref. \cite{Pengpra} will show an exponential decay with the atom number N, while our ECP can still obtain a relatively high success probability independent of N. Therefore, our ECP is efficient and may be useful and convenient in the current quantum information processing.

\section*{ACKNOWLEDGEMENTS}
 This work is supported by the National Natural Science Foundation of
China under Grant No. 11104159, Open Research
Fund Program of the State Key Laboratory of
Low-Dimensional Quantum Physics Scientific, Tsinghua University,
Open Research Fund Program of National Laboratory of Solid State Microstructures under Grant No. M25020 and M25022, Nanjing University,  and the Project Funded by the Priority Academic Program Development of Jiangsu
Higher Education Institutions.

\end{document}